# "I Felt Pressured to Give 100% All the Time": How Are Neurodivergent Professionals Being Included in SoPware Development Teams?

Nicoly da Silva Menezes
Universidade Federal do Pará Belém–Pa, Brazil
nicoly.menezes@icen.ufpa.br

Thayssa Águila da Rocha
Universidade Federal do Pará
Belém–Pa, Brazil
thayssa.rocha@icen.ufpa.br

Lucas Samuel Santiago Camelo
Universidade Federal do Pará Belém–Pa, Brazil lucas.camelo@itec.ufpa.br

Marcelle Pereira Mota
Universidade Federal do Pará
Belém–Pa, Brazil
mpmota@ufpa.br

## Abstract

**Context:** As the demand for digital solutions adapted to different user profiles increases, creating more inclusive and diverse software development teams becomes an important initiative to improve software product accessibility.

**Problem:** However, neurodivergent professionals are underrepresented in this area, encountering obstacles from difficulties in communication and collaboration to inadequate software tools, which directly impact their productivity and well-being.

**Solution:** This study seeks to understand the work experiences of neurodivergent professionals acting in different software development roles. A better understanding of their challenges and strategies to deal with them can collaborate to create more inclusive software development teams.

**IS Theory:** We applied the Sociotechnical Theory (STS) to investigate how the social structures of organizations and their respective work technologies influence the inclusion of these professionals.

**Method:** To address this study, we conducted semi-structured interviews with nine neurodivergent professionals in the Software Engineering field and analyzed the results by applying a continuous comparison coding strategy.

**Results:** The results highlighted issues faced by interviewees, the main ones related to difficulties in communication, social interactions, and prejudice related to their diagnosis. Additionally, excessive in work tools became a significant challenge, leading to constant distractions and cognitive overload. This scenario negatively impacts their concentration and overall performance.

**Contributions and Impact in the IS area:** As a contribution, this study presents empirically based recommendations to overcome sociotechnical challenges faced by neurodivergent individuals working in software development teams.

## 1 INTRODUCTION

The software development field has grown significantly in recent years, gaining visibility and attracting several professionals and students [18]. With the expansion in the creation of new products and, consequently, in the number of users, it is important to consider a wider diversity of profiles and different needs. In this context, it is important to consider the development teams' composition to reflect this plurality and support – from the initial stages – the process of building more inclusive and accessible products [1, 27].

Regardless, most work environments are not adapted to meet the needs of people with disabilities, both visible and hidden, as in some cases of neurodiversity [5]. Neurodiversity encompasses a variety of neurological conditions that affect human cognitive development. Among the most common are Attention Deficit Hyperactivity Disorder (ADHD), Autism Spectrum Disorder (ASD), and Dyslexia. In some psychology approaches, the term can also include other conditions that affect mental health, such as depression, and disorders that affect the learning process, such as dyscalculia [11].

Neurodivergent individuals have intrinsic characteristics in expressing themselves, rationalizing, and interacting in society [21]. People with ADHD, for example, tend to have a sharper sense of creativity and enthusiasm [16]. Attention to detail, systemic thinking, and good memory are characteristics of individuals with autism that can contribute positively to their performance in various areas of knowledge [3]. However, the job market continues to pose challenges to the inclusion of this group. Structural barriers contribute to the underrepresentation of neurodivergent people in various areas of activity, including Information Technology [3, 17]. The lack of understanding about the characteristics of these professionals, combined with social and cultural stigmas, often makes it difficult for them to enter and remain in the work market [21].

According to the social model of disability, it is society's responsibility to remove the barriers that prevent the full participation and inclusion of these individuals, ensuring their autonomy and contributions in various areas [23].

In this scenario, to guarantee the effective participation of neurodivergent professionals in the software engineering field, it is necessary to adopt some organizational strategies. Alternatives range from the creation of recruitment policies to value their skills, through practical adaptations in the workplace and implementing initiatives to raise awareness about different dimensions of diversity, thus fostering a more welcoming organizational culture [10, 31].

This study aims to explore the experiences of neurodivergent professionals working in different roles on software development teams in Brazil. To understand the challenges they face in the workplace, whether behavioral or tool-related, as well as the strategies to mitigate them, we developed two research questions:

(1) What are the main challenges faced by neurodivergent people working on software development teams?
(2) How can we modify the work environment, practices, and tools to support the inclusion of neurodivergent people on software development teams?

To achieve this goal, we conducted nine semi-structured interviews with nine software engineers diagnosed with Autism Spectrum Disorder (ASD), Attention Deficit Hyperactivity Disorder (ADHD), Obsessive-Compulsive Personality Disorder (OCPD), Central Auditory Processing Disorder (CAPD), and Obsessive-Compulsive Disorder (OCD).

Based on the lessons learned from these professionals' reports, we sought to expand knowledge and foster debate on including neurodivergent people in the software engineering field. This study's originality lies in addressing the gap in research focused on the integration of neurodivergent people within development teams, specifically in the context of the Brazilian software industry.

This paper is structured as follows: Section 2 presents the theoretical framework for this study. Section 3 discusses related studies that address similar themes. Section 4 details the research methodology, and the results are presented in Section 5. The discussion is raised in Section 6, and the final considerations and future works are presented in Section 7. Finally, Section 8 presents recommendations to support neurodivergent inclusion in development teams, followed by acknowledgments in Section 9.

## 2 THEORETICAL FRAMEWORK

To provide a better understanding of this work's context, we delimited our approach using the **Sociotechnical Theory**, which is based on the interrelationship between the social (human) and technical (tools) aspects in an organization. This perspective argues that organizational performance and success should not be explicitly restricted to the systems designed. However, it should also encompass the well-being of the individuals involved in the processes.

Thus, for an organization to truly thrive, it is important to balance technology and the employee's needs, skills, and relationships [7].

This theory emphasizes that a productive and sustainable work environment is achieved when social and technical elements are effectively integrated. In such an environment, employees feel valued and motivated to carry out their work. This, in turn, benefits the organization by strengthening its internal culture and enhancing employee commitment to their responsibilities [7].

Therefore, the theoretical framework presented incorporates social perspectives on neurodiversity and its technical implications specifically for the software engineering field.

### 2.1 Building the Concept of Neurodiversity

The theoretical thinking about neurodiversity began in the mid-1990s through discussions in communities composed exclusively of autistic children's parents or guardians. Over time, these communities experienced a process of emancipation and self-determination, being progressively occupied by autistic people themselves, resulting in the consolidation of this self-defense movement [19]. This movement was crucial in challenging stigmas and social prejudices related to autism. It promoted civil rights and recognized autism as a valid expression of individuality and a different way of being. [28, 29].

These discussions motivated the Australian sociologist Judy Singer to coin the term *neurodiversity* to designate individuals who present a variation in the development and functioning of their neurological structures, affecting their way of expressing, acting, and interacting. This perspective understands neurodiversity as a natural and inherent process of the human condition, rejecting the idea that there is a "standard" form of neural processing [29].

In Brazil, discussions about neurodiversity also held roots in activist movements in favor of the autistic community [24]. The Autistic Pride Brazil Movement (MOAB) [22], created in 2005, has played a fundamental role in the fight for civil rights, demystifying social stigmas related to the diagnosis of autism, and improving autistic people and their families quality of life. The Brazilian Law for the Inclusion of Persons with Disabilities, approved in 2015 [26], has propelled Brazil towards important advances in recognizing and understanding the needs of individuals with different conditions, including those with neurological diversity, in inclusion in educational and professional environments.

### 2.2 Software Engineering and Neurodiversity

Skills such as critical thinking, creativity, pattern analysis, and problem-solving are abilities cited in software engineering job descriptions. These skills, also known as "soft skills" [1], complement the technical knowledge required to perform everyday tasks such as code review, writing user stories, and identifying and correcting defects.

Neurodivergent people often have valuable characteristics for working in software engineering, such as the ability to focus deeply on complex tasks, as well as high attention to details and the ability to identify patterns and potential errors [3, 10, 20]. Companies that adopt inclusive practices, such as adaptations in physical spaces and flexibility in work dynamics, can benefit from the integration of these professionals in their teams due to the diversity of ideas and perspectives in the development of more inclusive solutions [25].

However, studies highlight specific challenges faced by neurodivergent people in the software engineering work market, which go beyond those encountered by neurotypical professionals, among them sensory overload and emotional detachment [13, 15]. Furthermore, problems related to interpersonal communication, interpreting teammates' emotions, distractions, and noisy environments can negatively contribute to their productivity and performance. These challenges require adaptation in work environments and understanding the different communication forms among team members [21].

Despite the advances and relevance of discussions about neurodiversity in recent years, the academic software engineering literature on this topic is still limited [27]. The prevalence of studies that mainly address the topic of autism [8] has left other conditions underrepresented in research. A broader approach, considering a diversity of conditions, can contribute to expanding discussions to other individuals on the neurodiversity spectrum.

---

[1]Set of behavioral and emotional skills developed to deal with everyday situations in the professional environment [6].

## 3 RELATED WORKS

We searched the ACM[2], IEEE[3], HICSS[4], and Emerald[5] databases to identify papers related on the topics of this study. We focused on studies that specifically explored the perspectives of neurodivergent individuals in the workplace and further narrowed our selection to those that concentrated on the context of software development.

To explore the experiences of people with ADHD working as software engineering professionals, the authors Liebel et al. [20] conducted a study to investigate the challenges and strategies of including neurodivergent people in software development teams. The study revealed challenges related to lack of attention, problems with personal organization, communication, and emotional regulation. The study's conclusions improve the debate about building more inclusive work environments for these professionals. However, the study does not explore other conditions related to neurodiversity.

Tomczak [30] explored the use of technology as a strategy to improve the performance of people with autism in the workplace. Based on interviews with professionals in the field of inclusive education and also some employers, the author identified adaptations in four main areas: effective communication, task organization, stress management, and sensory sensitivity. The proposed solutions included using emails and chats for communication, calendar apps and to-do lists for organization, and wearable devices for stress monitoring. The author indicates that adaptations can increase employability and improve the well-being of people with autism in the workplace. However, the study did not consider the context of software development or other neurological conditions.

In a study with 36 professionals diagnosed with different types of neurological conditions – such as ASD, ADHD, and Dyslexia – the authors Das et al. [10] explored how they performed their activities remotely during the COVID-19 pandemic and what adaptations were needed for the virtual work environment. The authors argued that remote working can offer greater flexibility for neurodivergent professionals. However, their inclusion depends on practices such as prior sharing of meeting presentations, meeting captioning and transcription, and communication tool adjustments to suppress other participants' noise.

To understand how research in the area of Human-Computer Interaction (HCI) has addressed neurodiversity in the context of the workplace, the authors Burtscher and Gerling [5] conducted a literature review on the subject, considering three types of neurological conditions: autism, ADHD, and dyslexia. The results showed a limitation of studies dealing with neurodivergence at work, highlighting that many existing works adopt a capacitist perspective to correct intrinsic characteristics of neurodivergent people, treating their needs as something to be adjusted or overcome. In addition, the self-determination of neurodiverse people was little explored in the selected studies. When addressed, this group's participation was limited to providing feedback on technologies already developed, excluding them from the initial development processes. The study did not explore the practical experiences of neurodiverse professionals in software engineering.

The authors Doyle and McDowall [12] investigated neurodiversity at work through a systematic review, looking at research in psychology, occupational therapy, and human resources. The research identified a significant lack of applied studies aimed at the inclusion of neurodivergent individuals in professional environments, as well as an unequal focus on autism to the detriment of other conditions, such as ADHD and dyslexia. It also highlights a need for more alignment between academic research and organizational practice, proposing a model for future research that considers interventions at individual and organizational levels. The study did not specifically explore the software engineering field.

On the other hand, the da Rocha et al. [9] systematic literature research particularly focused on studies that addressed people with disabilities (PWD) working as software development team members. Among the 39 papers identified, only seven addressed neurodivergent professionals, i.e., only 18% of the final sample, demonstrating the scarcity of studies on this subject.

## 4 METHOD

This study's nature is applied and exploratory. It seeks to understand the main challenges faced by neurodivergent professionals as members of software development teams. In addition to the challenges, practices and possible strategies or adaptations in work tools were identified to improve the productivity and inclusion of these professionals.

To achieve our objective, we conducted structured interviews with nine neurodivergent professionals working in different roles in Brazilian software development companies. The initial outreach to volunteer professionals was conducted via an online form, which collected general information from individuals expressing interest in participating in the study, as well as their contact details for the subsequent step. After this, semi-structured interviews were held with those volunteers who agreed to participate in the interview phase. The study received approval from the Human Research Ethics Committee (HREC), registered under the number 74659423.6.0000.0018. Completing the online form required the respondent's agreement to the informed consent form. Figure 1 presents the research steps.

To support open science, the form used in Step 1, the script used to guide the interviews, and their full transcripts are available (in Portuguese) in a public repository [6].

### 4.1 Data Collection

The online survey was shared through the researchers' social media platforms, including Instagram, LinkedIn, and WhatsApp, receiving a total of 39 responses between May to August 2024. The survey remains active to support ongoing research.

In this first step, all questions were mandatory and data were collected regarding age group, gender identity, racial self-declaration, region of residence, diagnosis, time since diagnosis was discovered, job position and, finally, a means of contact was requested so that we could proceed with scheduling and conducting the interviews.

---

[2]https://dl.acm.org/
[3]https://ieeexplore.ieee.org/Xplore/home.jsp
[4]https://shidler.hawaii.edu/itm/hicss
[5]https://www.emerald.com/insight/
[6]https://bit.ly/4kufCsT

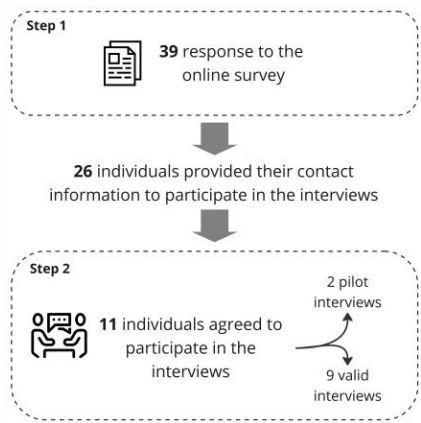

**Figure 1: Research Steps.**

Out of 39 responses received, only 26 respondents provided a means of contact (email or WhatsApp) to proceed to the next step. Furthermore, after the researchers contacted them, only 11 agreed to participate in the interview. To validate the proposed questions, two pilot interviews were conducted, and the remaining nine interviews were deemed valid for this study.

The interviews were conducted online using Google Meet from June to September 2024 and were recorded with the participants' consent, which was obtained orally. Each interview lasted between 20 and 60 minutes. The questions were designed to collect information about how participants shared their diagnoses in family and work contexts, as well as the challenges they encountered throughout their careers. Additionally, the discussions addressed the tools they used in their work, their experiences with these tools, and their preferences for work models. Participants were also asked about practices promoting inclusion in their workplace environments.

### 4.2 Data Analysis and Validation

The interview records were fully transcribed using the free transcription functionality available in the Microsoft Word tool[7]. Based on these transcriptions the open coding method was used to analyze the information. The method consists of a process of analysis, comparison, and categorization of qualitative data [14] that allows identification of the main recurring themes guided by the research questions defined.

The coding process was conducted by the first author and subsequently reviewed and discussed with the second and last authors to ensure the reliability and validity of the analysis. The coded representations were placed into Miro[8], an online visual collaboration and project management. These three authors participated in discussions about the mapped categories and codes, in order to reach a consensus on the interpretations of the data.

---
[7] https://support.microsoft.com/en-us/office/transcribe-your-recordings-7fc2efec-245e-45f0-b053-2a97531ecf57
[8] https://miro.com/

### 4.3 Researchers Profile

We are Brazilian researchers with expertise in Computer Science and Computer Engineering, focusing on Human-Computer Interaction (HCI) and accessibility. Our previous work examined the experiences of individuals with visual and auditory impairments within the software development area. In this study, we broaden our scope to encompass neurodiversity. We acknowledge the absence of an interdisciplinary approach as our academic backgrounds do not include specialization in Psychology or related fields. Our analysis explicitly examines this topic's practical implications in software engineering. We do not aim to make clinical judgments. As computer science researchers, we believe our role is to convey our participants' experiences and perceptions while respecting their narratives and supporting them in expressing their own identities.

## 5 RESULTS

The following subsections present the results obtained from the interviews. The participants' demographic data were obtained through the responses to the online survey. It is important to note that the translations of the interviewees' quotes aimed to faithfully capture their accounts, sometimes overlooking grammatical errors to preserve the integrity of their expressions.

### 5.1 Participant's Profiles

In order to outline the participants' profiles, we consolidated information considering aspects such as gender identity, race, region of residence, age range, position, diagnosis, and time since the discovery of the diagnosis. In general, the majority of the interviewees identified themselves as cisgender women (4 responses), declared themselves as black or brown (6 responses), and aged between 35 and 44 years old. The participants reported performing different roles in the area, such as test analyst, designer, software developer, and product manager.

Regarding diagnoses, the interviewees presented different neurological conditions, including ASD, ADHD, OCD, and OCPD. All participants reported having received the diagnosis in adulthood. Table 1 summarizes the participants' profile.

### 5.2 Challenges

The results presented here connect with our first research question: *What are the main challenges faced by neurodivergent people working on software development teams?*

*5.2.1 Lack of Family Support in Sharing Diagnosis.* This category was created from codes such as "stigma from family members", "neglect of diagnosis by parents", and "family skepticism". It addresses the difficulties faced by neurodivergent people in communicating their diagnosis even within the family context, especially due to the family's lack of understanding of the situation.

In total, three participants reported feeling fear or distrust when sharing their diagnoses with their relatives. In his testimony, P7 mentions that he does not intend to reveal his diagnosis due to his parent's failure to seek a diagnosis in childhood, in addition to previous conflicts related to his gender identity,

> P7: *"Yeah, I didn't say much to my relatives, like, I don't plan on talking, actually. [...] I'm trans first, right? And*

Table 1: Participant's Profile

| ID | Gender Identity | Age Group | Race | Most Recent Job Role | Diagnosis | Diagnosis Discovery |
|---|---|---|---|---|---|---|
| P1 | Cisgender woman | 25-34 years | White | Test Analyst | ADHD | < 5 years |
| P2 | Non-binary | 25-34 years | Brown | Software Developer | ADHD | < 3 years |
| P3 | Cisgender woman | 35-44 years | Brown | Project Coordinator | ADHD+CAPD | < 6 months |
| P4 | Cisgender man | 25-34 years | Brown | Designer | ASD | < 2 years |
| P5 | Cisgender man | 35-44 years | White | Software Developer | ASD+ADHD | < 3 years |
| P6 | Cisgender woman | 35-44 years | Brown | Product Manager | OCPD | < 2 years |
| P7 | Trans woman | 18-24 years | White | Test Analyst | ASD*+OCD | < 2 years |
| P8 | Cisgender woman | 18-24 years | Black | Software Developer | ASD+ADHD | < 2 years |
| P9 | Cisgender man | 35-44 years | Black | Software Developer | ASD+ADHD | < 6 years |

**Note:** *does not yet have a closed diagnosis for ASD due to a lack of financial resources to cover the tests.

*that already complicates life a bit and that's already created a bit of a rift. Because they're not the most open people to talking about it. And I'm afraid if I can talk about it. Anyway, they never went after the diagnosis."*

P1 and P2 reports also address their relatives' lack of credibility in their diagnoses obtained in adulthood. This behavior highlights a resistance to understanding neurodiversity in the family context. This resistance may be associated with socially constructed stigmas that hinder the understanding and acceptance of neuroatypical conditions. In addition, the failure to recognize symptoms in other family members reinforces a common pattern of neglect of initial manifestations within the family nucleus itself, silencing the expressions of neurodivergent people and preventing them from seeking support and adequate treatment.

P1: *"[...] my family doesn't accept it well, because they say 'Ah, you never had that, that's a fad, that's something made up'. But my mother, my brothers have several symptoms, but they didn't go after it, right?"*

P2: *"[...] they have a lot of that mentality that they are saturating neurodivergence diagnoses nowadays so they can push medication. [...] Neurodivergence runs in my family. I have an uncle who is on some level of the autism spectrum, and my father, he shows very strong signs that he is neurodivergent too, but he also doesn't pursue it because he doesn't believe in such a thing."*

*5.2.2 Negative Perceptions of Diagnosis in the Workplace.* The category was created from codes such as "fear of being blamed", "impact on selection processes", and "toxic organizational culture". This category details situations where participants reported feeling uncomfortable or afraid to share their diagnoses in the workplace due to biased colleague judgments and the risk of possible dismissal.

P1's report reflects the fear that neurodivergent people face when discussing their diagnoses in a professional context. In an organizational culture that disregards individual manifestations, distrust and hostility prevail. She worries about being judged by her colleagues, who may associate, in a prejudiced way, with eventual errors due to characteristics inherent to her diagnosis. This fear is intensified by the fear of suffering negative consequences, such as dismissal, for being transparent about her individualities.

P1: *"This new company I joined is a more traditional company, a company managed by older people and I didn't feel comfortable talking yet. They have a more archaic culture, so they have a culture of blaming many people. "Oh, it was so-and-so's mistake, it was so-and-so's mistake, it was so-and-so's mistake". And it's another reason for them to try to blame me for things related to my diagnosis [...] I think that if they knew the person was neurodivergent, they wouldn't even hire them."*

P4's report highlights an equally worrying aspect of the challenges and prejudices faced after sharing his diagnosis in the workplace. After receiving his autism diagnosis, P4 stated that he chose to be transparent and shared this information with his superiors. However, after that, he was fired. Since then, he has noticed that in the selection processes in which he mentioned his diagnosis, he did not receive any response from the recruiters. Meanwhile, he was contacted for the next steps in recruitment process when he omitted this information. This experience illustrates how neurodivergent professionals can face discrimination even before entering the job market, limiting their professional opportunities.

P4: *"In July I was diagnosed with autism. I shared the diagnosis with my employers and was fired, and since August until today I have participated in job selections and have not been able to get through. I got two interviews for positions where I did not share my diagnosis. But for the other positions where I shared my diagnosis I did not get any kind of callback or anything."*

*5.2.3 Emotional and Social Factors.* This category was created from codes such as "communication", "impulsivity", "harsh behavior" and "cognitive inflexibility".

The participants widely explored the social and emotional aspects, especially regarding interpersonal communication. Sometimes, because they have difficulty adapting their tone of communication, neurodivergent people are interpreted as harsh or rude. This perception is not always evident to them, becoming clearer when other people point out, as mentioned in P7's speech.

P7: *"But I think it's already happened, like the manager or boss talking about my communication, because sometimes I'm kind of harsh, rude and that can hurt people and everything. Like, in life I've been interpreted like that, being rude even when I don't mean to be."*

P4 also mentions communication difficulties. However, since he was not yet aware of his autism diagnosis, he sought out several conventional strategies in an attempt to overcome difficulties that he did not yet fully understand, investing in public speaking, non-violent communication, and other training. For P5, communication difficulties are a limiting aspect of his professional career, as he feels that the demands associated with this skill prevent him from reaching leadership positions. In addition, P5 describes his efforts to adapt to the behavior expected of a neurotypical employee.

> P4: *"Before I knew I was autistic, I had difficulty communicating with people, but I didn't know exactly what it was. So I took several public speaking courses, I took a nonviolent communication course, I took a group facilitation course, right? So I took several training courses to learn how to communicate."*

> P5: *"I try to act as if I were, as if I were an employee, let's say, a neurotypical employee. Only in situations where it's not possible, where it's beyond my capacity [...] I don't know if I'll ever have a management position, a leadership position, where I need to have communication skills, because that really is a challenge for me."*

P2's report brings up another aspect related to social interactions in the workplace, specifically regarding controlling impulsivity when communicating with others. For her, this impulsivity made it challenging to understand and actively listen to the opinions of other team members about strategies and decisions to be adopted. She highlights that therapy helped her improve this condition.

> P2: *"The main challenge I faced was relationships with people. The impulsiveness... of controlling impulsiveness, of saying that I think it has to be that way, without listening to people."*

Finally, the difficulty in making changes to scope very quickly, without carrying out proper planning, can cause mental exhaustion in these professionals, as mentioned by P4.

> P4: *"And also the issue of cognitive inflexibility, right? So, I have difficulty when there are very sudden changes in direction. I need some time to get my bearings. As startups, companies, they have this perspective that things change very quickly, I spend a lot more energy, I get much more exhausted adapting to these changes than my neurotypical peers."*

*5.2.4 Rigid Work Rules and High Workload.* This category was created from codes such as "pressure for productivity", "demand" and "lack of autonomy".

The lack of flexibility in working hours and other rigid work rules can negatively affect the productivity of neurodivergent professionals, generating demands from their superiors for not understanding their specificities. For P2, focus time is important for performing their activities more efficiently. However, this focus can occur at random times, including those outside the considered conventional, which led him to receive a warning from his superior.

> P2: *"So, for example, sometimes you'll find me working on a Saturday or at 8, 9 or 10 at night. But that's not because I'm working outside of work hours, it's because during work hours I just couldn't do that, I'd do something else. But sometimes, I've gotten yelled at for doing PR [pull requests] at 10 at night."*

P4 highlights the difference in perceptions about productivity. Quickly performing complex tasks is easy for him when he is absolutely focused. However, his leaders do not share this opinion and tend to associate his productivity only with long and intense workdays. These different perceptions about productivity cause frustration because he feels his contributions are less valued only because they do not fit the organization's standards.

> P4: *"So, often what would take me, I don't know, two or three days to do without focus, I can sometimes do in an afternoon with absolute focus. So, the number of hours worked, for me, is not synonymous with smart work or effort. Unfortunately, I realize that for leaders who are not involved in the daily activities, a person who works until 8 or 9 pm seems to be more committed than a person who starts on time and finishes on time [...] I felt pressured to give 100% all day every day, and that left me very exhausted."*

### 5.3 Adaptations in Practices

In this section, we present the results regarding possible work practices adaptations that may be beneficial in favoring inclusion, partially answering our second research question: *How can the work environment, its practices and tools, be modified to support the inclusion of neurodivergent people in software development teams?*

*5.3.1 Remote Work Model.* All interviewees unanimously indicated remote work as the most compatible with their demands and priorities when asked about the work model. In their statements, the participants mentioned some factors that positively impact mental health. An important point was feeling more motivated to carry out their activities in the comfort of their home, without worrying about the time spent commuting to the company office. P5 report highlights these points.

> P5: *"Well, the most obvious benefit is that I don't have to deal with increasingly chaotic traffic. It's two and a half hours that I gain from my day. If I were in the office right now, I would be stuck in traffic, arriving home maybe now. So these two hours that I gain represent an additional quality of life for me."*

Furthermore, the remote environment was considered more favorable for avoiding distractions or noise that could negatively impact the productivity of these professionals. On the other hand, the in-person environment can be seen as a threat to the mental health of neurodivergent individuals due to the stimuli generated through social interactions, which are quite common in this kind of environment. These interactions end up producing a high sensory load, affecting their well-being, as explained by P6 and P8.

> P6: *"[...] I've had to go to a physical location a few times, like meetings and such, and it's a very noisy environment, the social interactions, oh my   So I prefer remote because of that."*

> P8: *"I prefer remote work because first of all, in-person work requires a lot of sensory input, and secondly, a lot*

*of social interaction. So I'm constantly thinking about what I'm going to do and having to participate the way other people do."*

*5.3.2 Flexibility and Autonomy in the Workplace.* This category includes codes such as "flexibility in working hours", "freedom to slow down", and "learning time". In contrast to the problem highlighted in the previous subsection regarding rigid rules in the corporate environment, this category includes practices that allow adjustments in routine and work dynamics, making the environment less stressful for people with neurological diversity. For P8 and P9, for example, who face challenges related to learning time and understanding verbal commands, the support of their teammates is essential to alleviate possible frustrations related to intrinsic characteristics of their respective diagnoses.

> P8: *"I have the freedom to be slow in some things. And then when I joined (the team) I said 'look, my memory is terrible, I can't remember things, I'm going to have to ask a thousand times'. So I had people who stopped to explain the same thing to me 10 times until I understood, until I could do it on my own and, like, without feeling bad for it, or at least without letting it show they did. So that makes a big difference too."*

> P9: *"When I need to get information, I ask to record the call to review it several times until I understand it."*

Similarly, flexible working hours could benefit productivity by allowing participants, like P4, greater autonomy to manage their periods of intense focus on carrying out their activities. Reports demonstrate the importance of organizations adopting flexible work practices in order to promote a more inclusive and less stressful environment for these professionals.

> P4: *"So if I can have this freedom to organize my schedule and block it for a few hours a day so I won't be disturbed, that would be great."*

*5.3.3 Mental Health and Well-Being Support.* The category includes codes such as "encouragement of self-care" and "team support". This category addresses practices carried out by organizations that aim to promote their employee's well-being, taking into account the particularities of each individual.

Sensory stimuli, present in both remote and in-person work models, can significantly affect the performance of neurodivergent people in the workplace. Distractions, noise, or exposure to intense lighting can generate such a high sensory overload that it triggers high levels of stress, harming their mental health. This context can be particularly challenging for people with autism, as it can impact their engagement and full participation in team rituals. In attempts to overcome this adversity, P8 emphasizes the support of his colleagues, who prioritize his well-being and avoid pressuring him to attend meetings on days of greater sensitivity.

> P8: *"I know that it is okay if I get there and say 'I'm not feeling well, I'm not going to participate in the agenda anymore'. I need to make an effort to participate, but if it doesn't work out, that's okay too. Sometimes, something happens and I'm not feeling so well. And then I'm making an effort to be there and the guys say 'Take the day off, take the rest of the day off, rest, and you'll be back with the team tomorrow.'*

This report reinforces the value of a welcoming and respectful organizational culture that encourages self-care and creates favorable conditions for professionals to communicate freely about their impediments without feelings of guilt. For P5 and P3, the team's support has been essential in dealing with the demands of the work, providing the necessary support for learning gradually and collaboratively. This process involves an exchange of knowledge between the more experienced professionals who guide the younger ones.

> P5: *"The good thing is that in the team, there are people who help, more experienced people who pass on information to us little by little and we learn little by little, I think that's how it goes. [...] I've been lucky because I have colleagues willing to help me with this. So I can't help but mention that my colleagues have helped me a lot."*

Similarly, P3 shares a specific methodology adopted by his team where the less experienced professional learns by following all the activities performed by the more experienced coworker.

> P3: *"They have a methodology they call shadow, so you'll kind of be someone's shadow and you'll learn alongside that person, someone more experienced will be accompanying you. My manager once joined me on a call to help me set up an environment, configure my entire environment and that was really cool for me."*

*5.3.4 Training and Awareness Initiatives.* This category includes codes such as "lectures" and "training", covering some initiatives that companies are adopting to encourage debate around issues related to inclusion and diversity in the workplace. As noted by P3, lectures are held at the company where he works to promote respect for differences. These actions are highlighted as beneficial because they help create a welcoming and safe environment, benefiting both the organizational culture and the individuals within it.

> P3: *"Also, the HR people, they periodically give lectures on unconscious biases, talking about neurodivergence, talking about the LGBT community, talking about the importance of black people, brown people and everything else. So this ends up being a welcoming environment, I feel like the environment welcomes me."*

Some companies, as reported by P8, are adopting specific hiring processes for people with disabilities to increase their inclusion in development teams. These companies are also seeking strategies to retain these professionals in the organization through training rounds. People with autism, for example, often report difficulties in their communication, and to overcome this adversity, the company offers space to improve their behavioral skills.

> P8: *"I joined the company through a training process for people with disabilities. So, when I first joined, everyone already knew [...] So from time to time I had some calls regarding soft skills to understand what needed to change, which for me was not obvious, in terms of intonation, in the way of speaking."*

P9, in turn, highlights that there was training aimed at guiding his team in the appropriate reception and interaction with people with disabilities, reinforcing the importance of an inclusive approach in the workplace. As shown in the following excerpt.

> P9: *"My team was trained to be able to receive people with disabilities."*

Given these reports, it is possible to observe that these practices can benefit a broad group of people within an organization, including those with visible or hidden disabilities and other individuals belonging to minority groups in our society. These activities contribute to the understanding of social barriers that hinder the exercise of their basic rights, expanding the feeling of empathy and respect in the workplace.

*5.3.5 Time Management Methods.* The category includes codes such as "Time Blocking"[9] and "Pomodoro" and addresses the strategies adopted by neurodivergent professionals, specifically those with ADHD, to manage the time dedicated to developing their tasks. When asked about techniques to help with concentration on her activities, P2 mentions using the Time Blocking method to reduce cognitive overload when planning her actions during the week. On the other hand, P3 highlights the Pomodoro method to promote a deep focus on specific activities. This method helps avoid mental exhaustion and maintain productivity at consistent levels due to the alternation between intensive work and short rest intervals. These strategies allow neurodivergent to balance their work focus, reducing exhaustion and contributing to improving their productivity.

> P2: *"I've been trying several things for a long time now. What I've been using regularly for the last few months is Time Blocking, at the beginning of the week I separate all the big things I have to do that week."*

> P3: *"[...] but we helped each other that way, you know? By saying 'look, you have to write it down, use notes, you have to use the Pomodoro method', take a break for 10, 5, 15 minutes, and notice to see what your break time and concentration time is [...] with me I realized that it has to be 40 minutes, I can't go over 40 minutes."*

## 5.4 Adaptations in Work Tools

*5.4.1 Adjustments in Notification Management and Information Organization.* This category includes codes such as "notification overload", "excessive messaging", and "information organization". It covers the adaptations used by neurodivergent professionals to deal with the overload of notifications from the tools used and the difficulty in organizing the information passed on in the work environment. The participants' reports highlight the perception that frequent interruptions caused by notifications compromise their productivity and focus on developing their tasks.

In his report, P4 mentioned that when he chose to pause notifications to focus on his work activities, he was criticized for this behavior. This negative feedback generated a feeling of anxiety in him, highlighting the difficulty in balancing his need for focus with his superior's expectations of constant availability. P4's experience highlights the challenge that neurodivergent professionals face where the company's organizational culture does not accommodate the necessary adaptations so that they can act more productively and with less strain on their mental health.

> P4: *"[...] in the remote environment there was this issue of notifications which I usually silenced in order to focus on the activity. And I was criticized because I took a long time to respond to messages. So I was in a state of great anxiety when I had to keep the notifications turned on, because I couldn't focus 100%."*

The overload of notifications is mentioned again by P4 and also cited by P8, where both participants address the difficulty in managing this information, leading them to take action to avoid distractions and be able to focus on their activities.

> P4: *"What bothers me about Jira is [...] I receive too many emails from it, I can't organize myself in these emails, the notifications are also too many, so I need to do manual work of curating the information to know what is happening in each project."*

> P8: *" [...] when several people start sending messages to the group and they keep sending notifications all the time and I can't stop it. The only option I have is to mute it and it stays muted forever."*

*5.4.2 Use of Emerging Technologies to Improve Communication.* The category was created from the code "AI". This category highlights using artificial intelligence (AI) tools in the workplace to improve communication.

In an attempt to solve problems related to his considered harsh communication, P4 mentions using ChatGPT as a tool to mediate his communication in the corporate context. He uses it as a filter to soften messages, reduce misunderstandings, and help make his communication more pleasant for other team members. This report demonstrates how these technologies can serve as support strategies during the social interactions of neurodivergent professionals.

> P4: *"Sometimes I use it, I create a text-type filter, I pass it through GPT, I write my message and pass it through GPT to make this message more pleasant. And then this more pleasant message, is the one I share with people."*

## 6 DISCUSSION

Based on the results, we summarize the main findings regarding the research questions raised for this study. It was possible to observe that all interviewees went through the process of discovering their diagnosis only in adulthood, which generated significant challenges related to social acceptance and understanding of their identities throughout their personal and professional lives. These associated factors affect each individual's self-perception, often leading them to make great efforts to try to adapt to a neuronormativity pattern that does not support neurocognitive diversities, as highlighted in the literature by Bradley et al. [4].

Regarding the challenges, common barriers to neurodivergent individuals were found. The main obstacles are related to emotional and social factors, in addition to the inflexibility of conventional work standards, which do not always include necessary adaptations to prioritize mental health and optimize the productivity of these professionals. Furthermore, the difficulty in dealing with sudden changes, a recurring characteristic in technology companies, was also reported as a significant challenge [8]. Finally, the participants mentioned the fear of suffering adverse consequences regarding sharing the diagnosis in the workplace and recruitment processes [25]. These challenges can be addressed by adopting a more attentive and understanding perspective of organizations and valuing each individual's identity.

---
[9]A technique that divides daily tasks into specific time blocks [32].

Regarding inclusive practices, the reports mainly emphasized the importance of developing a welcoming and empathetic environment that understands professionals' different needs, rhythms, and work styles. In addition, training peers and other employees involved in the organization were cited as relevant for the adequate reception of neurodivergent workers. Adapting work dynamics and the importance of empathetic support from colleagues are also factors listed by da Rocha et al. [9] to promote a sense of belonging [2], in addition to increasing engagement in the activities performed. These practices can benefit a wide range of people within an organization, including individuals with visible or hidden disabilities and members of minority groups. Such initiatives raise awareness of the social barriers that obstruct the exercise of their basic rights, promoting a sense of empathy and respect in the workplace.

Customizing notifications in communication tools is important for supporting these individuals in their software development activities and reducing stress caused by constant interruptions. Organizations should provide resources that enable customization of functionalities, like allowing notifications to be muted during specific times. This flexibility allows adjust settings according to their needs, enhancing focus on tasks and organizing information more effectively. We also highlight the importance of establishing agreements when personalizing notifications, especially in full remote working teams. This resource is commonly used to establish the necessary communication among team members, and turning it completely off could harm the expected team collaboration.

Furthermore, emerging technologies such as artificial intelligence can be valuable tools to support neurodivergent individuals in their team interactions. However, these technologies should not be viewed as a means to correct characteristics that are intrinsic to neurodivergence. Companies should foster a culture of acceptance and respect for different forms of expression, striving to understand these characteristics rather than forcing them to conform to a perceived neuronormative standard.

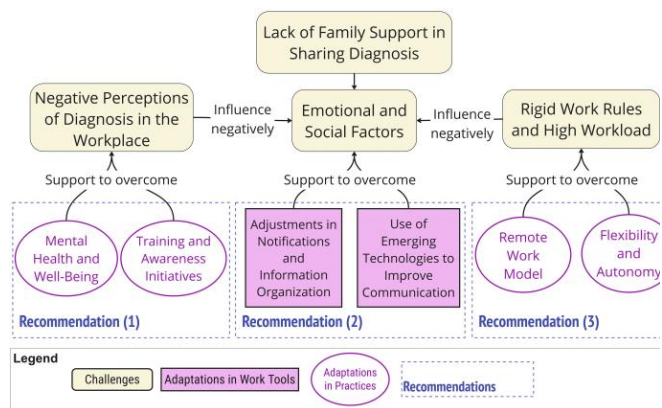

**Figure 2: Categories and Recommendations.**

## 7 RECOMMENDATIONS

Based on the results, it was possible to create a set of recommendations that companies can implement in their organizational structures to promote the well-being of neurodivergent professionals. Figure 2 presents the connection between the categories derived from the codes and their relationship to the recommendations.

(1) **Recognize and understand needs:** Neurodivergent people may have specific preferences for performing their activities. As a result, it is important to offer accommodations with low noise levels and controlled lighting. Managers should encourage these professionals to share their needs and preferences without fear of judgment or repercussions. In addition to accommodations, offering psychological support for these individuals can be equally important to make them comfortable in the workplace. Managers and their peers must be aware of their particularities, avoiding demands or expectations that lead to burnout, especially when forcing them to adapt to the neurotypical standard.

(2) **Prioritize and simplify communication channels:** Excessive tool notifications can cause stress and discomfort for neurodiverse people, negatively impacting their focus. To minimize these impacts, it is recommended to concentrate communications in a few channels and avoid sending numerous messages simultaneously and frequently. Moreover, fostering a culture of effective communication, where messages are clear and concise, can help lessen cognitive load on these professionals.

(3) **Supporting routine and schedule flexibility:** Neurodiverse people's focus may occur during specific periods, which do not always coincide with traditional business hours. Furthermore, allocating specific periods in one's schedule for focused work on particular activities could enhance the productivity of neurodivergent individuals. This flexibility allows for better time management and reduced distractions, which is important for improving productivity in tasks requiring high concentration levels. Therefore, an open dialogue with managers about adjusting the routine can result in a more productive and healthy work environment.

## 8 FINAL CONSIDERATIONS AND FUTURE WORK

This study explored the experiences of neurodivergent professionals who work in software development teams in Brazil, seeking to understand the main challenges they face, practices that benefit their inclusion, and possible adaptations in work tools. Semi-structured interviews were conducted with nine professionals with different neurological conditions, including ASD, ADHD, OCD, and OCPD.

This research contributes to the Information Systems area by providing a set of recommendations to support overcoming sociotechnical challenges faced by neurodivergent people in software development teams. These recommendations aim to inspire meaningful changes in organizational structures, making them more inclusive and adaptable. Additionally, they promote the personalization of work tools, ensuring they effectively address the needs of neurodivergent individuals and foster inclusion and well-being.

A limitation of this work is the number of participants involved in the study, which restricted the discussions to a specific group of neurodivergent individuals. Furthermore, no analyses were performed on other factors such as position, age group, or other dimensions of diversity, making it impossible broaden the debate on intersectionality in the neurodivergent community concerning gender or race, for example.

In future work, we intend to increase the number of participants, including individuals with other neurological diversities, to cover a more in-depth discussion considering other aspects related to the topic. We plan to explore the dynamics between neurodivergent professionals, their peers, and managers to identify how agreements that meet the organization's standards and the demands regarding these professionals' work environment adaptation needs can be implemented. Finally, we aim to explore proposals for interface adjustments in the main work tools used by software development teams to potentialize these individuals' professional performance in corporate environments.

## 9 ACKNOWLEDGEMENTS

Grammarly, a tool based on artificial intelligence, was used to support the writing and review of English grammar. Google Translate and DeepL, both tools for translation, were utilized to translate the original content into English. Finally, ChatGPT, an intelligence-based language model, was used to enhance text quality.